# A Multi-factor Adaptive Statistical Arbitrage Model

**Wenbin Zhang[1], Zhen Dai, Bindu Pan, and Milan Djabirov**
Tepper School of Business, Carnegie Mellon Unversity
55 Broad St, New York, NY 10005 USA

**Abstract**

This paper examines the implementation of a statistical arbitrage trading strategy based on co-integration relationships where we discover candidate portfolios using multiple factors rather than just price data. The portfolio selection methodologies include K-means clustering, graphical lasso and a combination of the two. Our results show that clustering appears to yield better candidate portfolios on average than naively using graphical lasso over the entire equity pool. A hybrid approach of using the combination of graphical lasso and clustering yields better results still. We also examine the effects of an adaptive approach during the trading period, by re-computing potential portfolios once to account for change in relationships with passage of time. However, the adaptive approach does not produce better results than the one without re-learning. Our results managed to pass the test for the presence of statistical arbitrage test at a statistically significant level. Additionally we were able to validate our findings over a separate dataset for formation and trading periods.

**Introduction**

Papers published in the past that explore co-integration and pairs trading identify portfolios of "similar" stocks by finding those whose prices historically moved in tandem. We felt that, in the co-integration case, this process can be improved upon by seeking "similar" stocks through measures other than price alone because the stock prices of characteristically similar firms will more or less move together. The intuition is that if we can identify portfolios that are alike over multiple dimensions, then their linear combinations (over price) should be more likely to revert to being co-integrated after any temporarily divergence. Injecting more information into the selection process by adding extra dimensions in order to identify stronger relationships in future price movements seemed worthwhile exploring. As a companion to graphical lasso, another machine learning technique - clustering was a natural choice to utilize. After briefly looking through published literature on co-integration, pairs trading, and other statistical arbitrage methodologies, we did not find any others attempting this concept.

The three major components for developing a statistical arbitrage are determining the right assets to trade, simulating trading through back testing, and verifying the existence of statistical arbitrage. Below is an outline of our study in these elements.

The first component, the selection process, highlights the bulk of our efforts:

- Factor selection: we used PCA technique to identify a set of independent factors. We used the factors themselves and the linear combination of these raw factors computed from PCA loadings.
- Clustering: we used K-mean clustering.

---
[1] Corresponding author. Email address: wenbinz@tepper.cmu.edu.

- Combining clustering and graphical lasso. We propose two distinct approaches – "Clustering-Glasso" and "Glasso-Clustering".

For the second component, we followed a standard strategy arbitrage trading procedure:

- We tested for a co-integration relationship for each identified portfolio.
- We checked whether the portfolio generated a positive profit over the formation period. If so, we continued to trade these portfolios.
- We attempted to rebalance the strategy during trading phase to account for clusters and co-integration relationships perhaps changing over time.

Finally, we used the JTTW-based approach to test the trading results and cross-validate our strategy.

**Data Collection and Normalization**

Our raw data was largely sourced from Bloomberg. We selected 19 different dimensions based on fundamental, statistical and momentum associated factors. This dataset covered all US stocks in the S&P 500 for the period starting from the first trading day of 2004 through the final trading day of 2011. The dimensions for our initial consideration are:

*Volatility (60 day)*
*Shares Outstanding*
*Sales Growth*
*RSI (Relative Strength Index)*
*Price to Book Ratio*
*Price to Sales Ratio*
*Price to EBITDA Ratio*
*P/E Ratio*
*Normalized ROE*
*Market Cap*
*Free Cash Flow Growth*
*Cash Flow Growth*
*Dividend (per share)*
*Bloomberg Estimates Analyst Rating*
*Total Number of Sell Recommendations*
*Total Number of Buy Recommendations*
*Price (close*
*Ask*
*Bid*

We cleaned the initial raw dataset by removing all non-trading days and missing values. There were 109 stocks with no missing values in all 19 dimensions across the entire period. Our implementation is based on this universe of stocks.

We note that it is probably more appropriate to have chosen the S&P 500 stocks from 2004 and enhanced our methodology to deal with missing fundamental data in separate formation periods. Unfortunately we did not manage to obtain the means to procure this data. This has the potential of introducing survivor bias. A separate section on data selection and potential bias re-visits this issue later in the paper.

Next, we normalized all dimensions before applying any additional filtering. The number of buy/sell recommendations were merged into a single factor as (buy-sell)/(buy+sell). We also took the logarithm of market cap and number of shares outstanding. This step is motivated Axtell who shows that US Firm sizes show a Zipf-law like distribution when plotted on a log-log scale (rank vs frequency). The factors were then normalized by subtracting the mean and dividing by the sample standard deviation.

Our date set will be divided into two parts:
- Regular Experiment Phase: From January 2004 to December 2007. The first two years are formation period, and the next two years are trading period.
- Cross Validation Phase: From January 2008 to December 2011. The first two years are formation period, and the next two years are trading period.

**PCA Analysis**

In order to select the factors that are most impactful we applied PCA over the normalized data. The below graphs shows the resulting analysis:

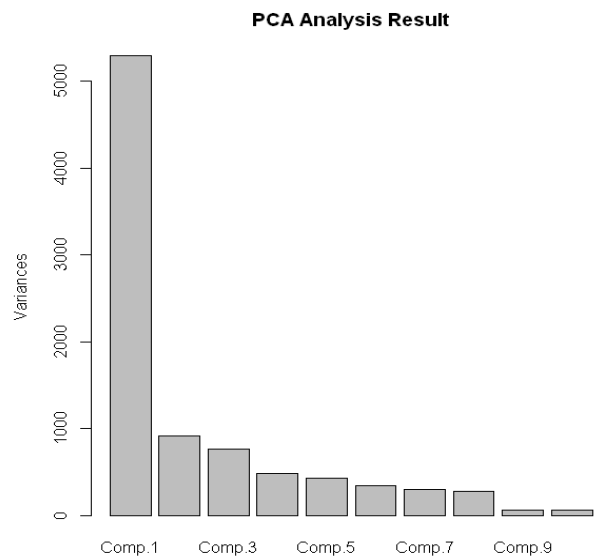

From the output of the loadings, we determined that the 7 most significant components contribute to 95.5% of the total variance. We used two different approaches towards factor selection given this data.

*Choosing Most Significant Raw Factors*

Based on the independent principal components generated by PCA, we can readily observe the dimensions that are largely responsible for variance of our data. In this case, we did not directly use the linear combinations. The 7 most significant factors are:

*P/E ratio*
*Price to Sales Ratio*
*Cash Flow Growth*
*Price*
*Price to EBITDA ratio*
*ROE*

*Volatility*

*Choosing Principal Components Generated by PCA*

We also directly chose the 7 most significant principal components for our analysis.

We ran clustering algorithms based on both selection approaches in the results to follow.

**K-mean Clustering**

There are a number of commonly used clustering algorithms. We felt, for our purpose, the most intuitive choice is K-means clustering. In order to produce a reasonable size for each cluster during the formation period, we chose K=30 which seems to generate cluster sizes of about 2-4 stocks on average.

**Candidate Portfolio Generation**

To keep the portfolio sizes comparable for each selection methodology, we enforced a policy of 2 - 4 stocks per portfolio. In this study, we applied two simple approaches (clustering and graphical lasso) and two hybrid approaches (Clustering-Glasso and Glasso-Clustering) to generate candidate trading portfolios.

*K-means Clustering*

- If a cluster contains only one stock, ignore.
- If a cluster contains 2, 3 or 4 stocks, take the entire cluster as a candidate portfolio.
- If a cluster contains 5 or more stocks, split them into sub-groups of 2 or 3 stocks and treat each group as a candidate portfolio.

For our initial formation period, this method generated 35 candidate trading portfolios with an average of 2.89 stocks per portfolio with selected 7 raw factors; and it generated 37 candidate trading portfolios with an average of 2.73 stocks per portfolio with top 7 principal components.

*Graphical Lasso (Glasso)*

- If there is only one non-zero entry in a given row of the inverse correlation matrix, ignore.
- If there are 2, 3 or 4 non-zero entries in a given row of the inverse correlation matrix, take the corresponding stocks as a candidate portfolio.
- If there are 5 or more non-zero entries in a given row of the inverse correlation matrix, take the corresponding 4 stocks with the largest absolute values as a candidate portfolio.

For our initial formation period, this method generated 55 candidate trading portfolios with an average of 3.82 stocks per portfolio.

*K-means Clustering - Graphical Lasso (Clustering-Glasso)*

- Run K-means with K = 3 to create 3 large clusters.
- Run graphical lasso on the entire set.
- If there is only one non-zero entry in a given row of the inverse correlation matrix, ignore.
- If there are 2, 3 or 4 non-zero entries in a given row of the inverse correlation matrix, check to make sure that they belong to the same cluster. If not, ignore.

- If there are 5 or more non-zero entries in a given row of the inverse correlation matrix, take the corresponding 4 stocks with the largest absolute values.

For our initial formation period, this method generated 49 candidate trading portfolios with an average of 3.61 stocks per portfolio with selected 7 raw factors; and it generated 50 candidate trading portfolios with an average of 3.7 stocks per portfolio with top 7 principal components.

Running K-means clustering first will generate at most 109 candidate portfolios since we determine 0 or 1 portfolios per row in the inverse correlation matrix.

*Graphical Lasso - K-means Clustering (Glasso-Clustering)*

- Run graphical lasso on the entire set.
- Run K-means with K = 3 to create 3 large clusters.
- Filter the inverse correlation matrix based on cluster membership, i.e. set up 3 separate passes through the inverse correlation matrix. When searching under one cluster, members of other clusters will have their entries in the inverse correlation matrix set to 0.
- For each pass, if there is only one non-zero entry in a given row of the inverse correlation matrix, ignore.
- If there are 2, 3 or 4 non-zero entries in a given row of the inverse correlation matrix, take the corresponding stocks as a candidate portfolio.
- If there are 5 or more non-zero entries in a given row of the inverse correlation matrix, take the corresponding 4 stocks with the largest absolute values as a candidate portfolio.

For our initial formation period, this method generated 132 candidate trading portfolios with an average of 3.53 stocks per portfolio with selected 7 raw factors. In this setup, each row of the inverse correlation matrix can produce up to 3 candidate portfolios, and as expected, given the methodology we chose, the number of candidate trading portfolios found increased significantly with this second attempt at a hybrid search approach. We thought that this second approach may have produced too many candidate portfolios. In fact we had significant amount of room to carry out additional selection and still have a comparable number of portfolios with respect to the other selection methods. To that end, we ranked each of the 132 portfolios by the sum of the absolute values of the non-zero entries in the inverse correlation matrix.

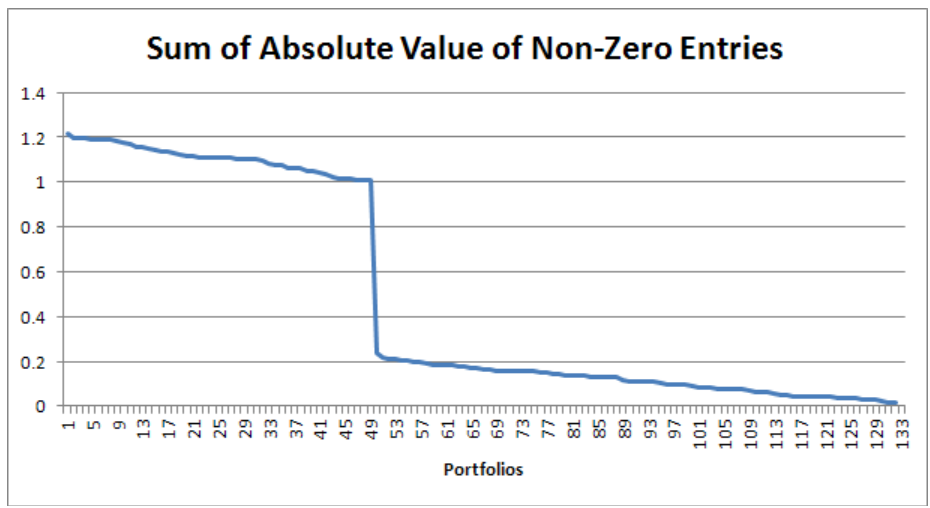

From this graph, we can see that 50 is an appropriate cut-off point to choose portfolios. In order to have a fair comparison, we choose 55 portfolios, the number detected by solely using the graphical lasso method, for our simulation on the next step.

**Portfolio Simulation**

We applied the standard Johansen test for co-integration relationship on the candidate portfolios determined by each selection method. Those portfolios that passed the test are experimentally traded over a formation period from January 2004 through December 2005. Those that produced a net positive profit in the formation period go on to be traded in the trading period from January 2006 through December 2007.

We normalized the long and short of our open trades such that the sum of their absolute values is $2. Below table shows the simulation result with portfolios based on solely clustering or graphical lasso method.

|  | **Clustering (Based on Sig. Raw Factors)** | | **Clustering (Based on Principal Components)** | | **Graphical Lasso** | |
| --- | --- | --- | --- | --- | --- | --- |
|  | Simulation Result | Remarks | Simulation Result | Remarks | Simulation Result | Remarks |
| Portfolios identified | 35 |  | 37 |  | 55 |  |
| Average # of stocks per portfolio | 2.89 |  | 2.73 |  | 3.82 |  |
| Portfolios passed Johansen test | 4 | 11.4%[1] | 6 | 16.2%[1] | 17 | 30.9%[1] |
| Portfolios that produce a net positive profit during formation period | 3 | 75%[2] | 5 | 83.3%[2] | 11 | 64.7%[2] |
| Portfolios that produce a net positive profit during trading period | 3 | 100%[3] | 3 | 60.0%[3] | 5 | 45.5%[3] |
| Total # of trades during trading period | 17 |  | 31 |  | 61 |  |
| Total # of trades that produce a net positive profit during trading period | 14 | 82.4%[4] | 26 | 83.9%[4] | 51 | 83.6%[4] |
| Average net profit per trade | 0.019 |  | 0.031 |  | 0.012 |  |
| Average net profit per portfolio | **0.109** |  | **0.194** |  | **0.067** |  |
| Total net profit | **0.327** |  | **0.97** |  | **0.737** |  |

[1] Ratio of portfolios passed Johansen test to total number of portfolios
[2] Ratio of portfolios generated a positive profit during formation period to portfolios passed Johansen test
[3] Ratio of portfolios generated a positive profit during trading period to portfolios generate a positive profit during formation period
[4] Ratio of trades produced a positive profit during trading period to all trades opened

We observed that the clustering algorithm identified fewer candidate portfolios. Additionally, percentage wise, a fewer of these portfolios passed the Johansen test. However, a greater percentage of them yielded a net positive profit in the trading period. The average net profit per trade and per portfolio is also significantly higher than that of the graphical lasso method.

Overall, clustering and graphical lasso yielded comparable performance in terms of generating candidate trading portfolios for co-integration-based statistical arbitrage strategy. Clustering found fewer portfolios but they were more profitable on average. We believe that the difference in the results come from the fact that clustering algorithms captures mainly cross-sectional behavior between stocks while graphical lasso concerns with only historical price time series.

Similarly we ran the same test for the two hybrid approaches with two different variable selection methods – most significant raw factors and principal components. In general, they all yielded higher profit per portfolio and higher total net profit, comparing to individual clustering or graphical lasso methods.

| Clustering based on Sig. Raw Factors (Sizes of three clusters: 32, 37, 40) | | | | |
|---|---|---|---|---|
| | **Clustering-Glasso** | | **Glasso-Clustering** | |
| | Simulation Result | Remarks | Simulation Result | Remarks |
| Portfolios identified | 49 | | 55 | |
| Average # of stocks per portfolio | 3.61 | | 3.62 | |
| Portfolios passed Johansen test | 18 | 36.7% | 19 | 34.6% |
| Portfolios that produce a net positive profit during formation period | 14 | 75% | 14 | 73.7% |
| Portfolios that produce a net positive profit during trading period | 11 | 77.8% | 11 | 77.8% |
| Total # of trades during trading period | 92 | | 83 | |
| Total # of trades that produce a net positive profit during trading period | 80 | 87.0% | 71 | 85.5% |
| Average net profit per trade | 0.032 | | 0.032 | |
| Average net profit per portfolio | **0.210** | | **0.190** | |
| Total net profit | **2.94** | | **2.66** | |

| Clustering based on Principal Components (Sizes of three clusters: 32, 35, 42) | | | | |
|---|---|---|---|---|
| | **Clustering-Glasso** | | **Glasso-Clustering** | |
| | Simulation Result | Remarks | Simulation Result | Remarks |
| Portfolios identified | 50 | | 55 | |
| Average # of stocks per portfolio | 3.7 | | 3.69 | |
| Portfolios passed Johansen test | 9 | 18.0% | 9 | 16.4% |
| Portfolios that produce a net positive profit during formation period | 8 | 88.9% | 8 | 88.9% |
| Portfolios that produce a net positive profit during trading period | 6 | 75.0% | 6 | 75.0% |
| Total # of trades during trading period | 43 | | 41 | |
| Total # of trades that produce a net positive profit during trading period | 36 | 83.7% | 34 | 82.9% |
| Average net profit per trade | 0.022 | | 0.027 | |
| Average net profit per portfolio | **0.121** | | **0.138** | |
| Total net profit | **1.09** | | **1.10** | |

We wanted to also make sure that our additional filtering in the graphical lasso-clustering method accurately sifted out less profitable candidates. The table below shows the simulation results from trading the top ranked 30/50/60/90/100 versus all 132 portfolios for the raw-factor clustering case. Indeed, we saw that the lowest ranked 22 portfolios did not add any value to the strategy.

| # of portfolios selected | 30 | 50 | 70 | 90 | 110 | 132 (All) |
|---|---|---|---|---|---|---|
| Average # of stocks per portfolio | 4.0 | 3.58 | 3.7 | 3.73 | 3.71 | 3.53 |
| Portfolios passed Johansen test | 11 | 19 | 24 | 29 | 33 | 34 |
| Portfolios that produce a net positive | 8 | 14 | 19 | 23 | 27 | 27 |
| profit during formation period | | | | | | |
| Portfolios that produce a net positive profit during trading period | 7 | 11 | 15 | 17 | 19 | 19 |
| Total # of trades during trading period | 48 | 83 | 114 | 140 | 158 | 158 |
| Total # of trades that produce a net positive profit during trading period | 41 | 71 | 97 | 119 | 133 | 133 |
| Average net profit per trade | 0.034 | 0.032 | 0.028 | 0.025 | 0.023 | 0.023 |
| Average net profit per portfolio | 0.201 | 0.190 | 0.166 | 0.154 | 0.134 | 0.134 |
| Total net profit | 1.608 | 2.66 | 3.15 | 3.54 | 3.618 | 3.618 |

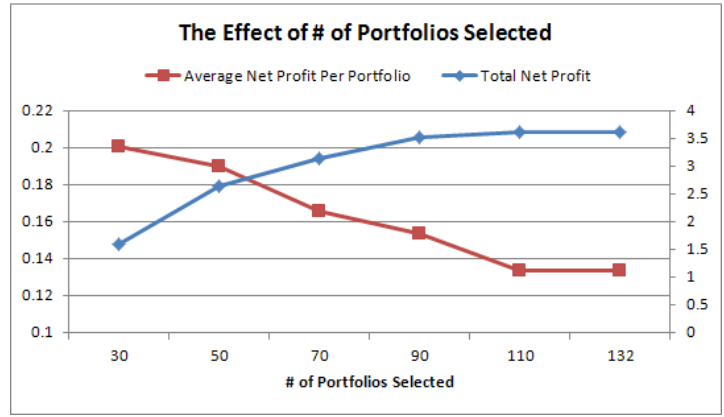

In addition, for the raw-factor-based clustering case, the histograms below show the profit distributions for clustering-glasso (49 portfolios), glasso-clustering (55 portfolios), and glasso-clustering (132 portfolios). We can see the center of the distribution plots is positive, though there is a somewhat longer tail on the negative side. We also observed very similar distribution plots for principal-components-based clustering.

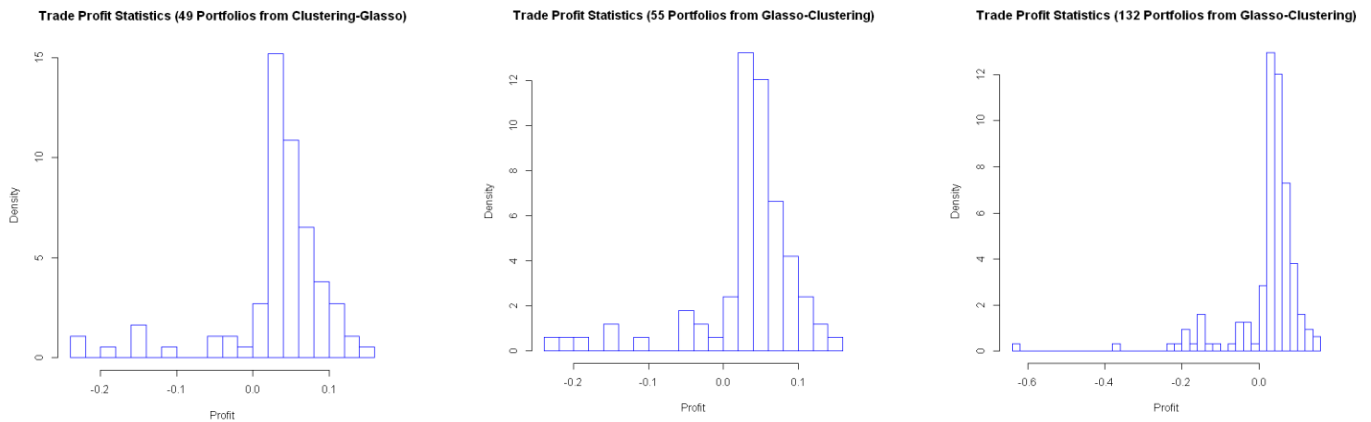

In summary, we observed that either hybrid yielded better results than clustering or graphical lasso alone, for both raw-factor-based clustering and principal-component-based clustering. The average net profit per trade and portfolio in both cases were raised significantly. Indeed using a combination of K-means clustering and graphical lasso casted a wide net over the possible candidate portfolios (comparable to using graphical lasso alone) as well as improved the overall selection quality. This improvement performance signaled that intrinsically, the two selection criteria likely did not overlap significantly. In addition, hybrid models opened more trades, which means they created more trading opportunities as well.

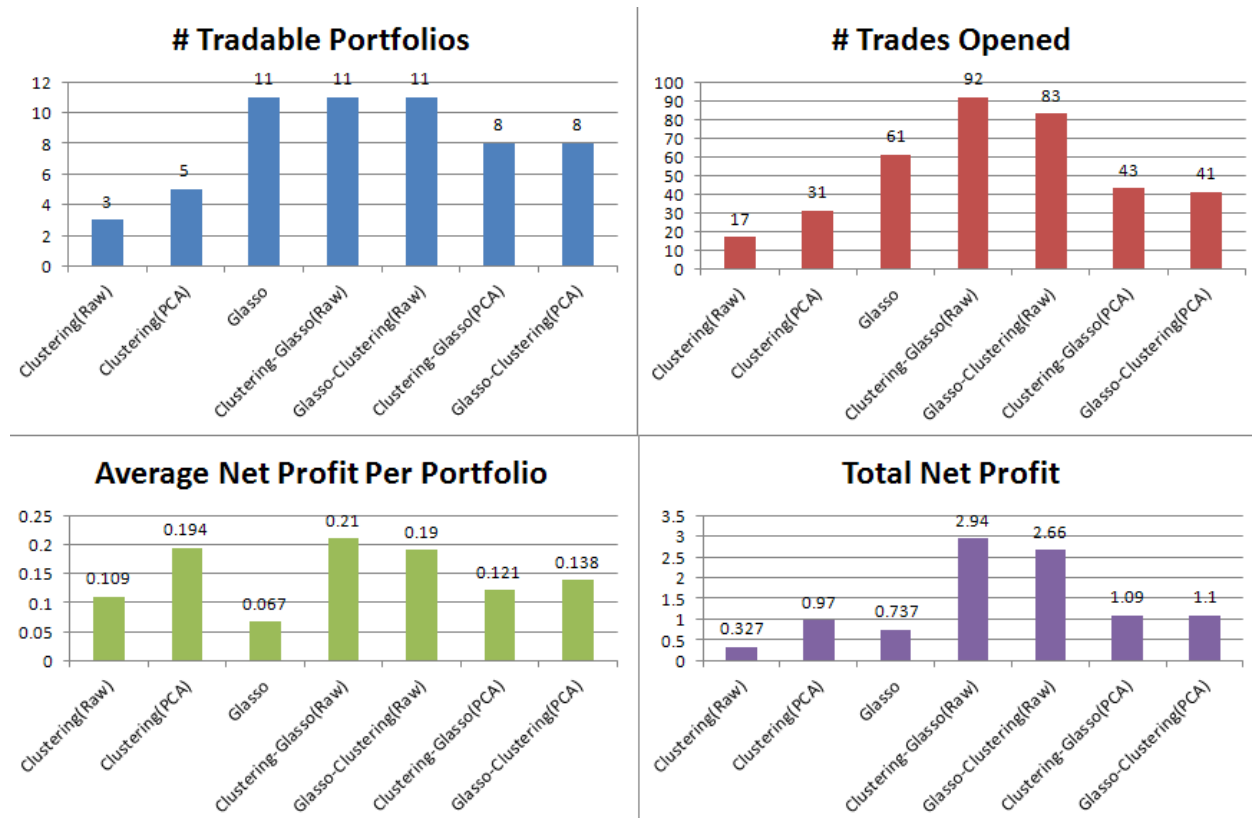

**Statistical Arbitrage Testing**

    We took two approaches to generating the P&L time series from our results for testing the existence of statistical arbitrage.
- In one, we applied a daily mark-to-mark approach to generating our gains and losses on our positions.
- In the other, we took the realized profit or loss on each trade and distributed the amount evenly, with discounting, and took daily average over the period of the holding.

    In both approaches, we fitted the JTTW model with an AR(1) noise term to each series. The time series for the risk free rate used was the daily 3 month Treasury bill rates from 2004 to 2011. From our experimental results, the realized P&L approach looked to be more informative because the trades opened did not evenly cover the entire trading period so that we saw a flat P&L series during certain time periods. We use a 0.05 significance level for all tests we performed.

    Under the singular portfolio selection methods, only principal-components-based clustering method passed our statistical arbitrage test, while the graphical lasso method and pure raw-factor-based clustering method did not pass the test. However, for both raw-factor-based clustering method and principal-

components-based clustering method, all two hybrid models (Clustering-Glasso and Glasso-Clustering) yielded very low p-values (<0.05), signaling that we should reject the null hypothesis that a statistical arbitrage does not exist. Therefore, our hybrid models produced statistical arbitrage strategies in all cases.

|  | **Clustering (Based on Sig. Raw Factors)** | | **Clustering (Based on Principal Components)** | | **Graphical Lasso** | |
|---|---|---|---|---|---|---|
|  | P-value | Remarks | P-value | Remarks | P-value | Remarks |
| Singular portfolio selection methods | 0.785 | Failed | 0.01 | Success | 0.234 | Failed |

|  | **Clustering-Glasso** | | **Glasso-Clustering** | |
|---|---|---|---|---|
|  | P-value | Remarks | P-value | Remarks |
| Clustering based on raw factors | 0.041* | Success | 0.0* | Success |
| Clustering based on principal components | 0.0* | Success | 0.0* | Success |

*\* All the hybrid models passed statistical arbitrage tests at a 0.05 significance level.*

**Adaptive Trading**

We tested rebalancing our portfolio once during the trading period by closing all trades at the end of 2006, re-running the two hybrid portfolio selection methods on 2006 data and trading the newly found candidates in 2007.

| **Clustering based on Sig. Raw Factors** <br> **(Sizes of three clusters: 32, 37, 40 in the first half, and 28, 58, 23 in the second half)** | | | | |
|---|---|---|---|---|
|  | **Clustering-Glasso** | | **Glasso-Clustering** | |
|  | Simulation Result | Remarks | Simulation Result | Remarks |
| Portfolios identified | 49/41 |  | 55/55 |  |
| Average # of stocks per portfolio | 3.61/3.39 |  | 3.62/3.53 |  |
| Portfolios passed Johansen test | 18/7 | 27.8% | 19/9 | 25.5% |
| Portfolios that produce a net positive profit during formation period | 14/2 | 64% | 14/2 | 57.1% |
| Portfolios that produce a net positive profit during trading period | 9/1 | 62.5% | 9/1 | 62.5% |
| Total # of trades during trading period | 66/7 |  | 58/7 |  |
| Total # of trades that produce a net positive profit during trading period | 55/5 | 82.2% | 46/5 | 78.5% |
| Average net profit per trade | 0.015/0.028 |  | 0.014/0.028 |  |
| Average net profit per portfolio | 0.071/0.098 |  | 0.057/0.098 |  |
| Total net profit | 0.994/0.196 |  | 0.798/0.196 |  |
| P-value of Statistical arbitrage test (Realized P&L) | 0.0/0.4 | Success | 0.0/0.4 | Success |

| **Clustering based on Principal Components** <br> **(Sizes of three clusters: 35, 32, 42 in the first half, and 26, 29, 54 in the second half)** | | | | |
|---|---|---|---|---|
|  | **Clustering-Glasso** | | **Glasso-Clustering** | |
|  | Simulation Result | Remarks | Simulation Result | Remarks |
| Portfolios identified | 50/39 |  | 55/55 |  |
| Average # of stocks per portfolio | 3.7/3.44 |  | 3.62/3.56 |  |
| Portfolios passed Johansen test | 9/5 | 15.7% | 11/7 | 16.4% |
| Portfolios that produce a net positive profit during formation period | 8/2 | 71.4% | 8/2 | 55.6% |
| Portfolios that produce a net positive profit during trading period | 6/1 | 70% | 8/1 | 90% |
| Total # of trades during trading period | 32/7 | | 36/7 | |
| Total # of trades that produce a net positive profit during trading period | 27/5 | 82.1% | 31/5 | 85.7% |
| Average net profit per trade | 0.029/0.028 | | 0.026/0.028 | |
| Average net profit per portfolio | 0.117/0.098 | | 0.119/0.098 | |
| Total net profit | 0.936/0.196 | | 0.952/0.196 | |
| P-value of Statistical arbitrage test (Realized P&L) | 0/0.03 | Success | 0/0.03 | Success |

This experiment still produced profitable trades on average throughout the trading period though less profitable than simply not rebalancing. We think this can largely contributed to forcibly closing out all trades at the end of 2006.

**Cross Validation**

Cross validation was performed on the second half of our cleaned data. The formation period was set from 2008 through 2009 and the trading period lasted from 2010 through 2011.

| | **Clustering (Based on Sig. Raw Factors)** | | **Clustering (Based on Principal Components)** | | **Graphical Lasso** | |
|---|---|---|---|---|---|---|
| | Simulation Result | Remarks | Simulation Result | Remarks | Simulation Result | Remarks |
| Portfolios identified | 34 | | 35 | | 90 | |
| Average # of stocks per portfolio | 2.88 | | 2.77 | | 3.87 | |
| Portfolios passed Johansen test | 13 | 38.2% | 9 | | 54 | 60% |
| Portfolios that produce a net positive profit during formation period | 9 | 69.2% | 6 | | 39 | 72.2% |
| Portfolios that produce a net positive profit during trading period | 6 | 66.7% | 5 | | 23 | 59.0% |
| Total # of trades during trading period | 64 | | 36 | | 194 | |
| Total # of trades that produce a net positive profit during trading period | 54 | 84.4% | 30 | | 160 | 82.5% |
| Average net profit per trade | 0.026 | | 0.017 | | 0.015 | |
| Average net profit per portfolio | **0.186** | | **0.103** | | **0.075** | |
| Total net profit | **1.674** | | **0.618** | | **2.925** | |
| P-value of Statistical arbitrage test (Realized P&L) | 0.0 | Success | 0.0 | Success | 0 | Success |

We saw that the results for clustering and graphical lasso alone are reasonably in line with what we saw in our initial testing. Actually clustering itself outperforms graphical lasso quite a bit.

Below two tables show the hybrid models with raw factors and principal components. The results consistently show that the hybrid models outperform sole clustering models or graphical lasso models.

| Clustering based on Sig. Raw Factors (Sizes of three clusters: 23, 29, 57) | | | | |
|---|---|---|---|---|
| | Clustering-Glasso | | Glasso-Clustering | |
| | Simulation Result | Remarks | Simulation Result | Remarks |
| Portfolios identified | 83 | | 90 | |
| Average # of stocks per portfolio | 3.77 | | 3.81 | |
| Portfolios passed Johansen test | 47 | 56.6% | 51 | 56.7% |
| Portfolios that produce a net positive profit during formation period | 39 | 83.0% | 40 | 78.4% |
| Portfolios that produce a net positive profit during trading period | 27 | 69.2% | 27 | 67.5% |
| Total # of trades during trading period | 208 | | 206 | |
| Total # of trades that produce a net positive profit during trading period | 173 | 83.2% | 169 | 82.0% |
| Average net profit per trade | 0.018 | | 0.016 | |
| Average net profit per portfolio | **0.096** | | **0.077** | |
| Total net profit | **3.744** | | **3.08** | |
| P-value of Statistical arbitrage test (Realized P&L) | 0.0 | Success | 0.0 | Success |

| Clustering based on Principal Components (Sizes of three clusters: 22, 41, 44) | | | | |
|---|---|---|---|---|
| | Clustering-Glasso | | Glasso-Clustering | |
| | Simulation Result | Remarks | Simulation Result | Remarks |
| Portfolios identified | 82 | | 90 | |
| Average # of stocks per portfolio | 3.84 | | 3.84 | |
| Portfolios passed Johansen test | 45 | 54.9% | 48 | 53.5% |
| Portfolios that produce a net positive profit during formation period | 30 | 66.7% | 33 | 68.9% |
| Portfolios that produce a net positive profit during trading period | 23 | 71.9% | 24 | 72.7% |
| Total # of trades during trading period | 154 | | 154 | |
| Total # of trades that produce a net positive profit during trading period | 127 | 82.5% | 125 | 81.2% |
| Average net profit per trade | 0.034 | | 0.030 | |
| Average net profit per portfolio | **0.174** | | **0.138** | |
| Total net profit | **5.22** | | **4.554** | |
| P-value of Statistical arbitrage test (Realized P&L) | 0.0 | Success | 0.0 | Success |

The raw-factor-based hybrid models performed a bit worse than the testing period. However, they still generated candidate portfolios that are more profitable than those detected by using the graphical lasso method alone. In particular, all hybrid models generated much higher total net profits than either clustering model or graphical model alone.

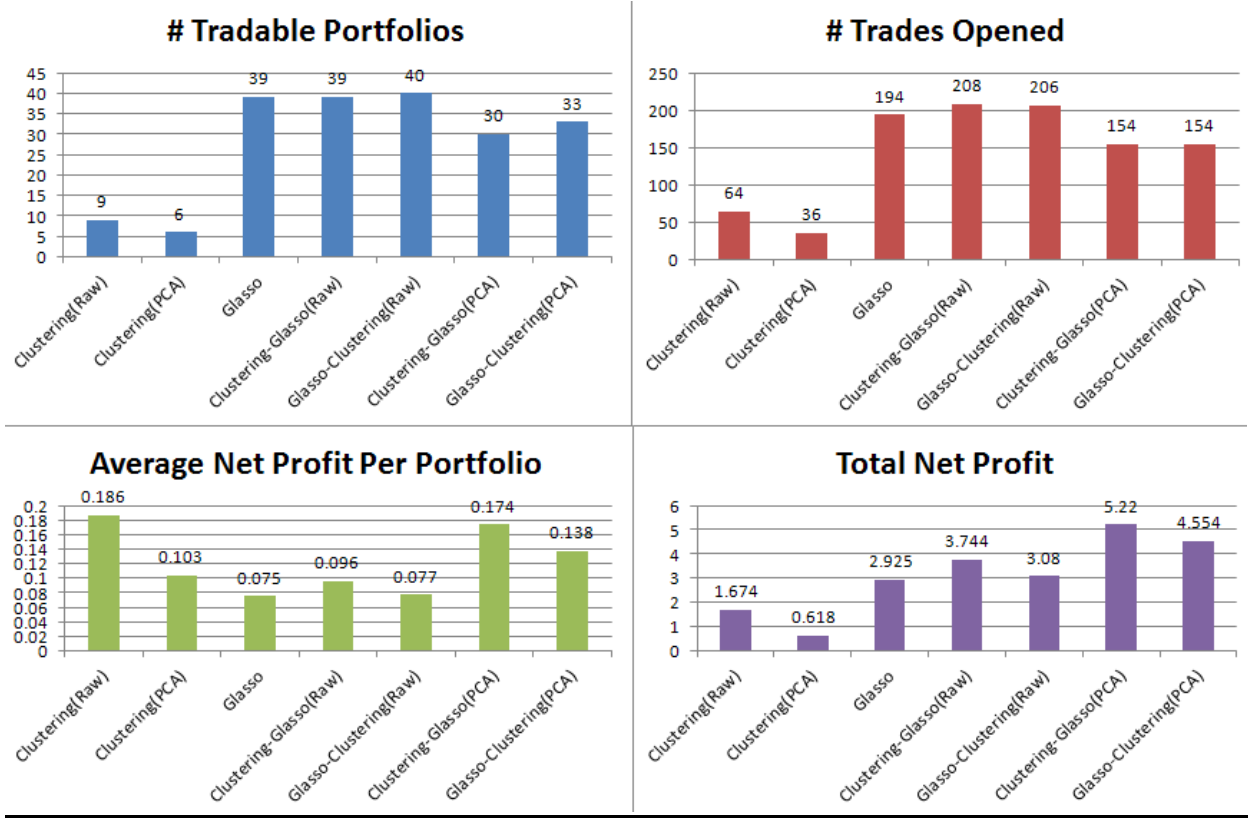

We also tested adaptive trading over the cross validation period. The results are shown below.

| Clustering based on Sig. Raw Factors (Sizes of three clusters: 23, 29, 57 in the first half, and 22, 32, 55 in the second half) | | | | |
|---|---|---|---|---|
| | Clustering-Glasso | | Glasso-Clustering | |
| | Simulation Result | Remarks | Simulation Result | Remarks |
| Portfolios identified | 83/86 | | 90/90 | |
| Average # of stocks per portfolio | 3.77/3.91 | | 3.81/3.87 | |
| Portfolios passed Johansen test | 47/30 | 45.6% | 51/29 | 44.4% |
| Portfolios that produce a net positive profit during formation period | 39/21 | 77.9% | 40/19 | 73.8% |
| Portfolios that produce a net positive profit during trading period | 23/10 | 55% | 23/10 | 55.9% |
| Total # of trades during trading period | 132/105 | | 128/101 | |
| Total # of trades that produce a net positive profit during trading period | 105/88 | 81.0% | 101/85 | 81.2% |
| Average net profit per trade | 0.009/0.011 | | 0.009/0.008 | |
| Average net profit per portfolio | 0.031/0.056 | | 0.028/0.041 | |
| Total net profit | 1.209/1.176 | | 1.12/0.779 | |
| P-value of Statistical arbitrage test (Realized P&L) | 0/0 | Success | 0.01/0 | Success |

| Clustering based on Principal Components (Sizes of three clusters: 24, 41, 44 in the first half, and 22, 29, 58 in the second half) | | | | |
|---|---|---|---|---|
| | Clustering-Glasso | | Glasso-Clustering | |
| | Simulation Result | Remarks | Simulation Result | Remarks |
| Portfolios identified | 82/86 | | 90/90 | |
| Average # of stocks per portfolio | 3.84/3.88 | | 3.84/3.83 | |
| Portfolios passed Johansen test | 45/32 | 45.8% | 48/26 | 41.1% |
| Portfolios that produce a net positive profit during formation period | 30/24 | 70.1% | 33/18 | 75.7% |
| Portfolios that produce a net positive profit during trading period | 16/14 | 55.6% | 16/12 | 50% |
| Total # of trades during trading period | 107/106 | | 108/88 | |
| Total # of trades that produce a net positive profit during trading period | 83/86 | 79.3% | 82/73 | 79.1% |
| Average net profit per trade | 0.010/0.013 | | 0.010/0.014 | |
| Average net profit per portfolio | 0.035/0.057 | | 0.032/0.067 | |
| Total net profit | 1/05/1.368 | | 1.056/1.206 | |
| P-value of Statistical arbitrage test (Realized P&L) | 0.01 | | 0.01 | |

We can see all the trade win ratios are quite high (around 80%), but the trading profits are lower than the non-adaptive case. Similar to what we saw during testing, we suspect that closing all positions at the end of 2010 negatively impacted our profitability because we may miss opportunities to gain profit on these trades in the near future. We did see some trades with very negative profits. (See below profit distribution chart.) We can see that the distribution is skewed. One solution is that we can set a lower bail-out threshold, for example 0.2 instead of 0.6. Our experiments show that the profit is improved greatly with this lower bail-out threshold.

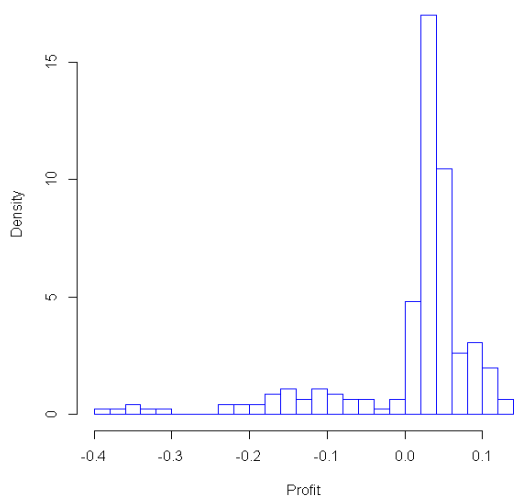

Trade Profit Statistics (Rebalancing of Glasso-Clustering)

## Survivorship Bias

One issue that needs special attention when analyzing our results is data selection and survivorship bias. We wanted to select a wide universe of stocks with readily available statistics on the 19 factors we used as input to our candidate portfolio selection strategy. A natural candidate was the SP500 index which is a widely recognized benchmark. Unfortunately obtaining historical compositions of SP500 proved difficult. While Standard and Poor's freely publishes current index composition, retrieving queries by date is part of a paid subscription service. Choosing the universe of stocks to be today's SP500 and not changing that when testing back in time already implies survivorship bias. However, while we cannot currently prove that, our belief is that year-over-year the index composition changes are small enough that the general validity of our results would still hold. To give an idea of how the SP500 changes over time, we mined the following data from various online news sources.

**SP500 Index Composition Changes:**

| 2007 | 2008 | 2009 | 2010 | 2011 |
| --- | --- | --- | --- | --- |
| TSS replaces SNV | SCG replaces MER | PLN replaces SGP | DISCA replaces PBG | WPX replaces CPWR |
| WPO replaces TIN | FLIR replaces NCC | FSLR replaces WYE | BRK.B replaces BNI | TEL replaces CEPH |
| RRC replaces TRB | OI replaces WB | ARG replaces CBE | KMX replaces XTO | MOS replaces NSM |
| GME replaces DJ | MFE replaces BRL | CFN replaces MTW | QEP replaces STR | MPC replaces MRO |
| AMT replaces AT | EQT replaces RIG | FMC replaces CTX | ACE replaces MIL | AMB replaces PLD |
| MTW replaces TEK | RSG replaces AW | RHT replaces CIT | TYC replaces SII | ANR replaces MEE |
| POM replaces HCR | DNB replaces LIZ | PWR replaces IR | IR replaces PTV | CMG replaces NOVL |
| TIE replaces BOL | LIFE replaces ABI | WDC replaces EQ | CVC replaces KG | BLK replaces GENZ |
| JEC replaces AV | SRCL replaces BUD | PCS added | FFIV replaces NYT | JOY replaces AYE |
| | CEPH replaces GGP | TEL deleted | NFLX replaces ODP | |
| | | | NFX replaces EK | |

There is an average of 10 or so ticker changes (out of 500) per year or around 2% of ticker turnover, which over a few years is probably not enough to introduce significant changes in the results. While ideally we would want to have time-specific composition of SP500 and account for missing data during the trading, we still believe our results hold valid especially when used in a comparative setting (glasso vs clustering and hybrid approaches) where all strategies face the same data.

## Conclusions

Based on our study, we felt that there is certainly merit in refining the portfolio selection process when developing a co-integration trading strategy. While a standalone graphical lasso approach detected a large number of candidate portfolios in our universe of stocks, their average profitability was relatively low. In contrast, a clustering only approach found fewer, but more profitable candidate portfolios. A hybrid approach was able to benefit from the strength of both by generation a reasonable number of profitable portfolios. We were not able to find a similar level of result in our implementation of continuously rebalancing portfolios, but we feel that there is room for improvement on this front.

## Future Work

As mentioned earlier, given ticker histories, we could have gathered more data, for example the equities in the S&P 500 in 2004 rather than 2012. This would have fully eliminated any potential for look-ahead bias and survivorship bias. We can easily account for stocks that stop trading in our system during the trading period but our data selection actually ensures existence so such provisions would not trigger. Gathering the data with missing tickers turned out to be quite difficult. While we were not able to directly compare the universe of stocks from S&P 500 in 2004 versus that of 2012, we do not believe that universe was markedly different based on the data in more recent years. Moreover, given that few stocks were selected relative to the size of the universe, we do not believe that there is a strong presence of survivorship bias in our study, and we do believe our hybrid models are still able to beat clustering or graphical model alone consistently, but we need to re-verify this anyway once we obtain the "unbiased" dataset in our future research.

In terms of the stock selection process, we also wanted to experiment with other machine learning concepts such as hierarchical clustering or K-nearest neighbor classifier. Among partition-based clustering algorithms we could attempt applying fuzzy C-means clustering as well. Regarding the adaptive trading phase of our study, we would try to see the results of not forcibly closing trades at the end of 2006 and instead only update our pool of candidate portfolios for future holdings.

Additionally, we can tweak parameters more carefully in each step of our study, and we can apply systematical and adaptive approach to stop loss under highly risky environment. Actually we saw during the cross validation phase that there were a few trades that closed with large losses. From the distribution of profits we can see that a lower bail out threshold, e.g. 0.2 may have been more appropriately. Indeed when we made this adjustment, we saw marked improved in the average profit of each traded portfolio.